\documentclass[sigconf]{acmart}
\acmConference[ICSE 2022]{The 44th International Conference on Software Engineering}{May 21--29, 2022}{Pittsburgh, PA, USA}
\acmBooktitle{ICSE '22: The 44th International Conference on Software Engineering, May 21--29, 2022, Pittsburgh, PA, USA}
\acmPrice{15.00}
\acmISBN{978-1-4503-XXXX-X/18/06}

\usepackage[scale=0.99]{FiraMono}
\usepackage{caption}
\usepackage{subcaption}
\usepackage{amsmath,amsthm}
\usepackage{booktabs} 
\usepackage{graphicx}
\usepackage{colortbl}
\usepackage{pifont}
\usepackage{tabu}
\usepackage{color,soul}
\usepackage{textgreek}
\usepackage{proof}
\usepackage{upgreek}
\usepackage{extarrows}
\usepackage{MnSymbol}
\usepackage{forest}
\usepackage{multirow}
\usepackage{marvosym}
\usepackage{listings}
\usepackage{pgfpages}
\usepackage{tikz}
\usepackage{listings}          
\usepackage{parcolumns}
\usepackage{etoolbox}
\usepackage{hyperref}
\usepackage{epstopdf}
\usepackage{algorithm}
\usepackage{algorithmicx}
\usepackage{enumitem}
\usepackage[noend]{algpseudocode}

\setlist{nolistsep,leftmargin=*}

\algrenewcommand{\algorithmiccomment}[1]{{\hfill \textcolor{gray}{$\triangleright$ #1}}}
\theoremstyle{definition}
\newtheorem{definition}{Definition}

\theoremstyle{remark}

\newcommand{\code}[1]{{\ttfamily \small #1}}
\newcommand{\scode}[1]{{\footnotesize \ttfamily #1}}

\definecolor{mono1}{RGB}{56,58,66}
\definecolor{mono2}{RGB}{105,108,119}
\definecolor{mono3}{RGB}{156,158,171}

\definecolor{cyan}{RGB}{1,132,188}
\definecolor{blue}{RGB}{64,120,242}
\definecolor{purple}{RGB}{166,38,164}
\definecolor{green}{RGB}{80,161,79}

\definecolor{red1}{RGB}{228,86,73}
\definecolor{red2}{RGB}{202,18,67}

\definecolor{orange1}{RGB}{152,104,1}
\definecolor{orange2}{RGB}{193,132,1}

\definecolor{hlyellow}{RGB}{255,251,219}

\newcommand\digitstyle{\color{orange1}}

\makeatletter
\newcommand{\ProcessDigit}[1]
{%
  \ifnum\lst@mode=\lst@Pmode\relax%
   {\digitstyle #1}%
  \else
    #1%
  \fi
}
\lstdefinelanguage{Python}{
	basicstyle=\scriptsize\ttfamily\color{mono1},
	classoffset=0,
	morekeywords={def, await, import, lambda, break, while, in, return, else, pass, for, del, from, global, not, class, as, elif, if, yield, assert},
	keywordstyle=\color{purple!70!black},
	classoffset=1,
	morekeywords={random},
	keywordstyle=\color{red1},
	classoffset=2,
	morekeywords={True, False, None},
	keywordstyle=\color{orange1},
	classoffset=3,
	morekeywords={print, type, str, float, Exception,dict, int,  raise, list,sum},
	keywordstyle=\color{cyan!80!black},
	classoffset=4,
	morekeywords={round,remove, Custom,MyObj, get_betraying_probability, choice, range, read_csv},
	keywordstyle=\color{cyan},
	identifierstyle=\color{mono1},
	sensitive=false,
	comment=[l]{},
	stringstyle=\color{green}\ttfamily,
	commentstyle=\itshape\color{mono1!80}\ttfamily,
	numberstyle=\color{mono3}\ttfamily,
}

\usepackage{tcolorbox}
\newcounter{findingCounter}
\newenvironment{finding}{
\begin{tcolorbox}[colback=blue!5!white,colframe=blue!5!white,arc=0mm,grow to left by=0mm,left=0mm,grow to right by=0mm,left=1.5mm,right=1.5mm,top=1.5mm,bottom=1.5mm]
\textbf{Finding \arabic{findingCounter}\stepcounter{findingCounter}:}
}
{
\end{tcolorbox}
}

\newcommand{\name}{Nalin} 

\begin{document}

\title[Learning from Runtime Behavior to Find Name-Value Inconsistencies]{
\name{}: Learning from Runtime Behavior to Find\\ Name-Value Inconsistencies in Jupyter Notebooks
}

\author{Jibesh Patra}
\affiliation{
  \institution{University of Stuttgart}
  \country{Germany}
}
\email{jibesh.patra@gmail.com}

\author{Michael Pradel}
\affiliation{
  \institution{University of Stuttgart}
  \country{Germany}
}
\email{michael@binaervarianz.de}

\begin{abstract}

Variable names are important to understand and maintain code.
If a variable name and the value stored in the variable do not match, then the program suffers from a \emph{name-value inconsistency}, which is due to one of two situations that developers may want to fix:
Either a correct value is referred to through a misleading name, which negatively affects code understandability and maintainability, or the correct name is bound to a wrong value, which may cause unexpected runtime behavior.
Finding name-value inconsistencies is hard because it requires an understanding of the meaning of names and knowledge about the values assigned to a variable at runtime.
This paper presents \name{}, a technique to automatically detect name-value inconsistencies.
The approach combines a dynamic analysis that tracks assignments of values to names with a neural machine learning model that predicts whether a name and a value fit together.
To the best of our knowledge, this is the first work to formulate the problem of finding coding issues as a classification problem over names and runtime values.
We apply \name{} to 106,652 real-world Python programs, where meaningful names are particularly important due to the absence of statically declared types.
Our results show that the classifier detects name-value inconsistencies with high accuracy, 
that the warnings reported by \name{} have a precision of 80\% and a recall of 76\% w.r.t.\ a ground truth created in a user study,
and that our approach complements existing techniques for finding coding issues.

\end{abstract}




\maketitle

\section{Introduction}

Variable names are a means to convey the intended semantics of code.
Because meaningful names are crucial for the understandability and maintainability of code~\cite{Butler2010}, developers generally try to name a variable according to the value(s) it refers to.
Names are particularly relevant in dynamically typed languages, e.g., Python and JavaScript, where the lack of types forces developers to rely on names, e.g., to understand what types of values a variable stores.

Unfortunately, the name and the value of a variables sometimes do not match, which we refer to as a \emph{name-value inconsistency}.
A common reason is a \emph{misleading name} that is bound to a correct value.
Because such names make code unnecessarily hard to understand and maintain, developers may want to replace them with more meaningful names.
Another possible reason is that a meaningful name refers to an \emph{incorrect value}.
Because such values may propagate through the program and cause unexpected behavior, developers should fix the corresponding code.


The following illustrates the problem with two motivating examples, both found during our evaluation on real-world Python code~\cite{10.1145/3173574.3173606}.
As an example of a misleading name consider the following code:
\vspace{-.1em}
\begin{lstlisting}
log_file = glob.glob('/var/www/some_file.csv')
\end{lstlisting}
\vspace{-.1em}
The right-hand side of the assignment yields a list of file names, which is inconsistent with the name of the variable it gets assigned to, because \code{log\_file} suggests a single file name.
The code is even more confusing since this specific call to \code{glob} will return a list with at most one file name. That is, a cursory reader of the code may incorrectly assume this file name to be stored in the \code{log\_file} variable, whereas it is actually wrapped into a list.
To clarify the meaning of the variable, it could be named, e.g., \code{log\_files} or \code{log\_file\_list}, or the developer could adapt the right-hand side of the assignment by retrieving the first (and only) element from the list.
We find misleading names to be the most common reason for name-value inconsistencies.

Less common, but perhaps even worse, are name-value inconsistencies caused by an incorrect value, as in the following example:
\vspace{-.5em}
\begin{lstlisting}
train_size = 0.9 * iris.data.shape[0]
test_size = iris.data.shape[0] - train_size
train_data = data[0:train_size]
\end{lstlisting}
\vspace{.5em}
The code tries to divide a dataset into training and test sets.
Names like \code{train\_size} are usually bound to non-negative integer values.
However, the above code assigns the value 135.0 to the \code{train\_size} variable, i.e., a floating point value.
Unfortunately, this value causes the code to crash at the third line, where \code{train\_size} is used as an index to slice the dataset, but indices for slicing must be integers.
While the root cause and the manifestation of the crash are close to each other in this simple example, in general, incorrect values may propagate through a program and cause hard to understand misbehavior.

Finding name-value inconsistencies is difficult because it requires both understanding the meaning of names and realizing that a value that occurs at runtime does not match the usual meaning of a name.
As a result, name-value inconsistencies so far are found mostly during some manual activity.
For example, a developer may point out a misleading name during code review, or a developer may stumble across an incorrect value during debugging.
Because developer time is precious, tool support for finding name-value inconsistencies is highly desirable.

This paper presents \underline{\name{}}, an approach for detecting \underline{na}me-va\underline{l}ue \underline{in}consistencies automatically.
The approach combines dynamic program analysis with deep learning.
At first, a dynamic analysis keeps track of assignments during an execution and gathers pairs of names and values the names are bound to.
Then, a neural model predicts whether a name and a value fit together.
When the dynamic analysis observes a name-value pair that the neural model predicts to not fit together, then the approach reports a warning about a likely name-value inconsistency.

While simple at its core, realizing the \name{} idea involves four key challenges:
\begin{itemize}[leftmargin=1.7em]
\item[C1] Understanding the semantics of names and how developers typically use them.
The approach addresses this challenge through a learned token embedding that represents semantic similarities of names in a vector space.
For example, the embedding maps the names \code{train\_size}, \code{size}, and \code{len} to similar vectors, as they refer to similar concepts.
\item[C2] Understanding the meaning of values and how developers typically use them.
The approach addresses this challenge by recording runtime values and by encoding them into a vector representation based on several properties of values.
The properties include a string representation of the value, its type, and type-specific features, such as the shape of multi-dimensional numeric values.
\item[C3] Pinpointing unusual name-value pairs.
We formulate this problem as a binary classification task and train a neural model that predicts whether a name and a value match.
To the best of our knowledge, this work is the first to detect coding issues through neural classification over names and runtime values.
\item[C4] Obtaining a dataset for training an effective model.
The approach addresses this challenge by considering observed name-value pairs as correct examples, and by creating incorrect examples by combining names and values through a statistical, type-guided sampling that is likely to yield an incorrect pair.
\end{itemize}

Our work relates to learning-based bug detectors~\cite{oopsla2018-DeepBugs,Allamanis2018b,Li2019,Dinella2020,Wang2020a}, which share the idea to classify code as correct or incorrect.
However, we are the first to focus on name-value inconsistencies, whereas prior work targets other kinds of problems.
\name{} also relates to learned models that predict missing identifier names~\cite{Raychev2015,Context2Name,David2020}.
Our work differs by analyzing code with names supposed to be meaningful, instead of targeting obfuscated or compiled code.
Finally, there are static analysis-based approaches for finding inconsistent method names~\cite{Nguyen2020,Liu2019,Host2009} and other naming issues~\cite{He2021}.
A key difference to all the above work is that \name{} is based on dynamic instead of static analysis, allowing it to learn from runtime values, which static analysis can only approximate.
One of the few existing approaches that learn from runtime behavior~\cite{Wang2020} aims at finding vector representations for larger pieces of code, but cannot pinpoint name-value inconsistencies.

We train \name{} on 780k name-value pairs and evaluate it on 10k previously unseen pairs from real-world Python code extracted from Jupyter notebooks.
The model effectively distinguishes consistent from inconsistent examples, with an F1 score of 0.89.
Comparing the classifications by \name{} to a ground truth gathered in a study with eleven developers shows that the reported inconsistencies have a precision of 80\% and a recall of 76\%.
Most of the inconsistencies detected in real-world code are due to misleading names, but there also are some inconsistencies caused by incorrect values.
Finally, we show that the approach complements state-of-the-art static analysis-based tools that warn about frequently made mistakes, type-related issues, and name-related bugs.


In summary, this paper contributes the following:
\begin{itemize}
	\item An automatic technique to detect name-value inconsistencies.
    \item The first approach to find coding issues through neural machine learning on names and runtime behavior.
    \item A type-guided generation of negative examples that improves upon a purely random approach. 
    \item Empirical evidence that the approach effectively identifies name-value pairs that developers perceive as detrimental to the understandability and maintainability of the code.
\end{itemize}

\section{Overview}
\begin{figure*}[t]
	\includegraphics[width=\linewidth]{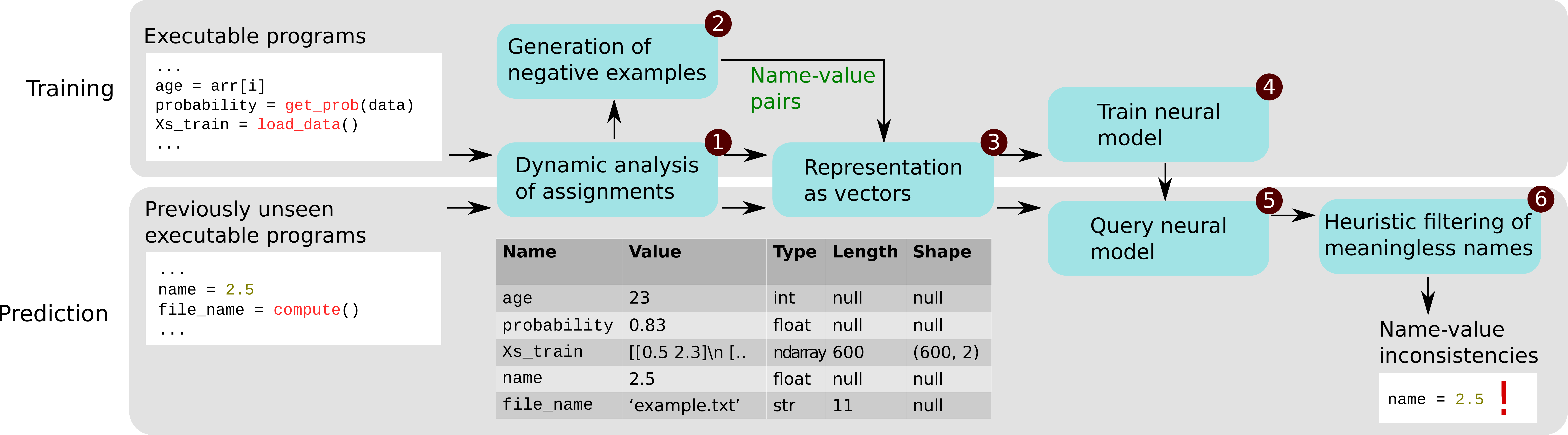}
	\caption{Overview of the approach.}
	\label{fig:overview}
\end{figure*}

This section describes the problem we address and gives an overview of our approach.
\name{} reasons about \emph{name-value pairs}, i.e., pairs of a variable name and a value that gets assigned to the variable.
The problem we address is to identify name-value pairs where the name is not a good fit for the value, which we call \emph{inconsistent name-value pairs}.
Identifying such pairs is an inherently fuzzy problem:
Whether a name fits a value depends on the conventions that programmers follow when naming variables.
The fuzziness of the problem motivates a data-driven approach~\cite{NeuralSoftwareAnalysis}, where we use the vast amounts of available programs as guidance for what name-value pairs are common and what name-value pairs stand out as inconsistent.

Broadly speaking, \name{} consists of six components and two phases, illustrated in Figure~\ref{fig:overview}.
During the training phase, the approach learns from a corpus of executable programs a neural classification model, which then serves during the prediction phase for identifying name-value inconsistencies in previously unseen programs.
The following illustrates the six components of the approach with some examples.
A detailed description follows in Section~\ref{sec:approach}.

Given a corpus of executable programs, the first component is a dynamic analysis of assignments of values to variables.
For each assignment during the execution of the program, the analysis extracts the variable name, the value assigned to the variable, and several properties of the value, e.g., the type, length, and shape.
As illustrated in Figure~\ref{fig:overview}, properties that do not exist for a particular value are represented by \emph{null}.
For example, the analysis extracts the length of the assigned value for \code{Xs\_train}, but not for \code{age} and \code{probability}, as the corresponding values are primitives that do not have a length. 

While the name-value pairs obtained by the dynamic analysis serve as positive examples, the second component generates negative examples that combine names and values in an unusual and likely inconsistent way.
The motivation behind generating negative examples is that \name{} trains a classification model in a supervised manner, i.e., the approach requires examples of both consistent and inconsistent name-value pairs.
Using the example pairs in Figure~\ref{fig:overview}, one negative example would be the name \code{Xs\_train} paired with the floating point value 0.83, which indeed is an unusual name-value pair.
Our approach for generating negative examples is a probabilistic algorithm that biases the selection of unusual values toward unusual types based on the types of values that are usually observed with a name.
The first and second component together address challenge~C4 from the introduction, i.e., obtaining a dataset for training an effective model.

The third component of \name{} addresses challenges~C1 and~C2, i.e., ``understanding'' the semantics of names and values.
To this end, the approach represents names and values as vectors that preserve their meaning.
To represent identifier names, we build on learned token embeddings~\cite{Bojanowski2017}, which map each name into a vector while preserving the semantic similarities of names~\cite{icse2021}.
For example, the vector of \code{probability} will be close to the vectors of names \code{probab} and \code{likelihood}, because these names refer to similar concepts.
To represent values, we present a neural encoding of values based on their string representation, type, and other properties.

Given the vector representations of name-value pairs, the fourth component trains a neural model to distinguish positive from negative examples.
The result is a classifier that, once trained with sufficiently many examples, addresses challenge~C3.
The fifth component of the approach queries the classifier with vector representations of name-value pairs extracted from previously unseen programs, producing a set of pairs predicted to be inconsistent.
The final component heuristically filters pairs that are likely false positives, and then reports the remaining pairs as warnings to the developer.
For the two new assignments shown in Figure~\ref{fig:overview}, the trained classifier will correctly identify the assignment \code{name = 2.5} as unusual and raises a warning.

\section{Approach}
\label{sec:approach}

The following presents the components of \name{} outlined in the previous section in more detail. 

\subsection{Dynamic Analysis of Assignments}
\label{subsec:data_extraction}

The goal of this component is to gather name-value pairs from a corpus of programs.
Our analysis focuses on assignments because they associate a value with the name of a variable.
One option would be to statically analyze all assignments in a program.
However, a static analysis could capture only those values where the right-hand side of an assignment is a literal, but would miss many other assignments, e.g., when the right-hand side is a complex expression or function call.
In the code corpus used in our evaluation, we find that 90\% of all assignments have a value other than a primitive literal on the right-hand side, i.e., a static analysis could not gather name-value pairs from them.
Instead, \name{} uses a dynamic analysis that observes all assignments during the execution of a program.
Besides the benefit of capturing assignments that are hard to reason about statically, a dynamic analysis can easily extract additional properties of values, such as the length or shape, which we find to be useful for training an effective model.

\subsubsection{Instrumentation and Data Gathering}

To dynamically analyze assignments, \name{} instruments and then executes the programs in the corpus.
For instrumentation, the analysis traverses the abstract syntax tree of a program and augments all assignments to a variable with a call to a function that records the name of the variable and the assigned value.

As runtime values can be arbitrarily complex, the analysis can extract only limited information about a value.
We extract four properties of each value, which we found to be useful for training an effective model, but extending the approach to gather additional properties of values is straightforward.
Slightly abusing the term ``pair'' to include the properties extracted for each value, the analysis extracts the following information:
\begin{definition}[Name-value pair]
    A name-value pair is a tuple $(n, v, \tau, l, s)$, where $n$ denotes the variable name on the left hand side, $v$ is a string representation of the value, $\tau$ represents the type of the value, and $l$ and $s$ represent the length and shape of the value, respectively.
\end{definition}
The string representation builds upon Python's built-in string conversion, which often yields a meaningful representation because developers commonly use this representation, e.g., for debugging.
The type of values is relevant because it allows \name{} to find type-related mistakes, which otherwise remain easily unnoticed in a dynamically typed language.
Length here refers to the number of items present in a collection or sequence type value, which is useful, e.g., to enable the model to distinguish empty from non-empty collections.
Since some common data types are multidimensional the \emph{shape} refers to the number of items present in each dimension.
The table in Figure~\ref{fig:overview} shows examples of name-value pairs gathered by the analysis.
We show in the evaluation how much the extracted properties contribute to the overall effectiveness of the model.

%
%
%

\subsubsection{Filtering and Processing of Name-Value Pairs}
\label{sec:data filtering}




\paragraph{Merge Types}
We observe that the gathered data forms a long-tailed distribution of types.
One of the reasons is the presence of many similar types, such as Python's dictionary type \emph{dict} and its subclass \emph{defaultdict}.
To help the model generalize across similar types, we reduce the overall number of types by merging some of the less frequent types.
To this end, we first choose the ten most frequent types present in the dataset.
For the remaining types, we replace any types that are in a subclass relationship with one of the frequent types by the frequent type.
For example, consider a name-value pair \textit{(stopwords, frozenset(\{"all", "afterwards", "eleven", ...\}), frozenset, 337, null)}.
Because type \emph{frozenset} is not among the ten most frequent types, but type \emph{set} is, we change the name-value pair into \textit{(stopwords, frozenset(\{"all", "afterwards", "eleven", ...\}), set, 337, null)}.

\paragraph{Filter Meaningless Names}
An underlying assumption of \name{} is that developers use meaningful variable names.
Unfortunately, some names are rather cryptic, such as variables called \code{a} or \code{ts\_pd}.
Such names help neither our model nor developers in deciding whether a name fits the value it refers to, and hence, we filter likely meaningless names.
The first type of filtering considers the length of the variable names and discards any name-value pairs where the name is less than three characters long.
The second type of filtering is similar to the first one, except that it targets names composed of multiple subtokens, such as \code{ts\_pd}.
We split names at underscores\footnote{\url{https://www.python.org/dev/peps/pep-0008/\#function-and-variable-names}}, and remove any name-value pairs where each subtoken has less than three characters.

\subsection{Generation of Negative Examples}
\label{subsec:create_negative_examples}

The gathered name-value pairs provide numerous examples of names and values that developers typically combine.
\name{} uses supervised learning to train a classification model that distinguishes consistent, or positive, name-value pairs from inconsistent, or negative, pairs.
Based on the common assumption that most code is correct, we consider the name-value pairs extracted from executions as positive examples.
The following presents two techniques for generating negative examples.
First, we explain a purely random technique, followed by a type-guided technique that we find to yield a more effective training dataset.

\subsubsection{Purely Random Generation}
Our purely random algorithm for generating negative examples is straightforward.
For each name-value pair  $(n, v, \tau, l, s)$, the algorithm randomly selects another name-value pair $(n', v', \tau', l', s')$ from the dataset.
Then, the algorithm creates a new negative example by combining the name of the original pair and the value of the randomly selected pair, which yields $(n, v', \tau', l', s')$.

While simple, the purely random generation of negative examples suffers from the problem of creating many name-value pairs that do fit well together.
The underlying root cause is that the distribution of values and types is long-tailed, i.e., the dataset contains many examples of similar values among the most common types.
For example, consider a name-value pair gathered from an assignment \code{num = 23}.
When creating a negative example, the purely random algorithm may choose a value gathered from another assignment \code{age = 3}.
As both values are positive integers, they both fit the name \code{num}, i.e., the supposedly negative example actually is a legitimate name-value pair.
Having many such legitimate, negative examples in the training data makes it difficult for a classifier to discriminate between consistent and inconsistent name-value pairs.

\subsubsection{Type-Guided Generation}
\label{sec:type-guided}

 \begin{algorithm}[tb]
    \caption{Create a negative example}
        \label{alg:create_a_negative_example}
        \begin{algorithmic}[1]
            \Require Name-value pair $(n, v, \tau, l, s)$, dataset $D$ of all pairs
            \Ensure Negative example $(n, v', \tau', l', s')$
            \State $F_{\mathit{global}} \leftarrow$ Compute from $D$ a map from types to their frequency
            \State $F_{\mathit{name}} \leftarrow$ Compute from $D$ and $n$ a map from types observed with $n$ to their frequency
            \State $T_{\mathit{name}} \leftarrow \emptyset$ \Comment{Types seen with $n$} \label{line:typesets start}
            \State $T_{\mathit{name\_infreq}} \leftarrow \emptyset$ \Comment{Types infrequently seen with $n$}
            \ForAll {($\overline{\tau} \mapsto f) \in F_{\mathit{name}}$}
                \State $T_{\mathit{name}} \leftarrow \overline{\tau}$ 
                \If{$f \leq$ threshold}
                    \State $T_{\mathit{name\_infreq}} \leftarrow \overline{\tau}$
                \EndIf
            \EndFor \label{line:typesets end}
            \State $T_{\mathit{all}} \leftarrow \mathit{dom}(F_{\mathit{global}})$ \Comment{All types ever seen} \label{line:neg value start}
            \State $T_{\mathit{cand}} = (T_{\mathit{all}} \setminus T_{\mathit{name}}) \bigcup T_{\mathit{name\_infreq}}$ \Comment{Types never or infrequently seen with $n$}
            \State $\tau' \leftarrow \mathit{weightedRandomChoice}(T_{\mathit{cand}}, F_{\mathit{global}})$
            \State $v', l', s' \leftarrow \mathit{randomChoice}(D, \tau')$ \label{line:pick val}
            \State \Return $(n, v', \tau', l', s')$ \label{line:neg value end}
        \end{algorithmic}
\end{algorithm}

To mitigate the problem of legitimate, negative examples that the purely random generation algorithm suffers from, we present a type-guided algorithm for creating negative examples.
The basic idea is to first select a type that a name is infrequently observed with, and to then select a random value among those observed with the selected type.
Algorithm~\ref{alg:create_a_negative_example} shows the type-guided technique for creating a negative example for a given name-value pair.
The inputs to the algorithm are a name-value pair $(n, v, \tau, l, s)$ and
the complete dataset $D$ of positive name-value pairs.

The first two lines of Algorithm~\ref{alg:create_a_negative_example} create two helper maps, which map types to their frequency.
The $F_{\mathit{global}}$ map assigns each type to its frequency across the entire dataset $D$, whereas the $F_{\mathit{name}}$ map assigns each type to how often it occurs with the name $n$ of the positive example.
Next, lines~\ref{line:typesets start} to~\ref{line:typesets end} populate two sets of types.
The first set, $T_{\mathit{name}}$, is populated with all types ever observed with name $n$.
The second set, $T_{\mathit{name\_infreq}}$, is populated with all types that are infrequently observed with name $n$.
``Infrequent'' here means that the frequency of the type among all name-value tuples with name $n$ is below some threshold, which is 3\% in the evaluation.
The goal of selecting types that are infrequent for a particular name is to create negative examples that are unusual, and hence, likely to be inconsistent.

The remainder of the algorithm (lines~\ref{line:neg value start} to~\ref{line:neg value end}) picks a type to be used for the negative example and then creates a negative name-value pair by combining the name $n$ with a value of that type.
To this end, the algorithm computes all candidates types, $T_{\mathit{cand}}$, that are either never observed with name $n$ or among the types $T_{\mathit{name\_infreq}}$ that infrequently occur with $n$.
The algorithm then randomly selects among the candidate types, using the global type frequency as weights for the random selection.
The rationale is to choose a type that is unlikely for the name $n$, while following the overall distribution of types.
The latter is necessary to prevent the model from simply learning to spot unlikely types, but to instead learn to find unlikely combinations of names and values.
Once the target type $\tau'$ for the negative example is selected, the algorithm randomly picks a value among all values (line~\ref{line:pick val}) observed with type $\tau'$, and eventually returns a negative example that combines name $n$ with the selected value.

\begin{figure}
    \includegraphics[width=.95\linewidth]{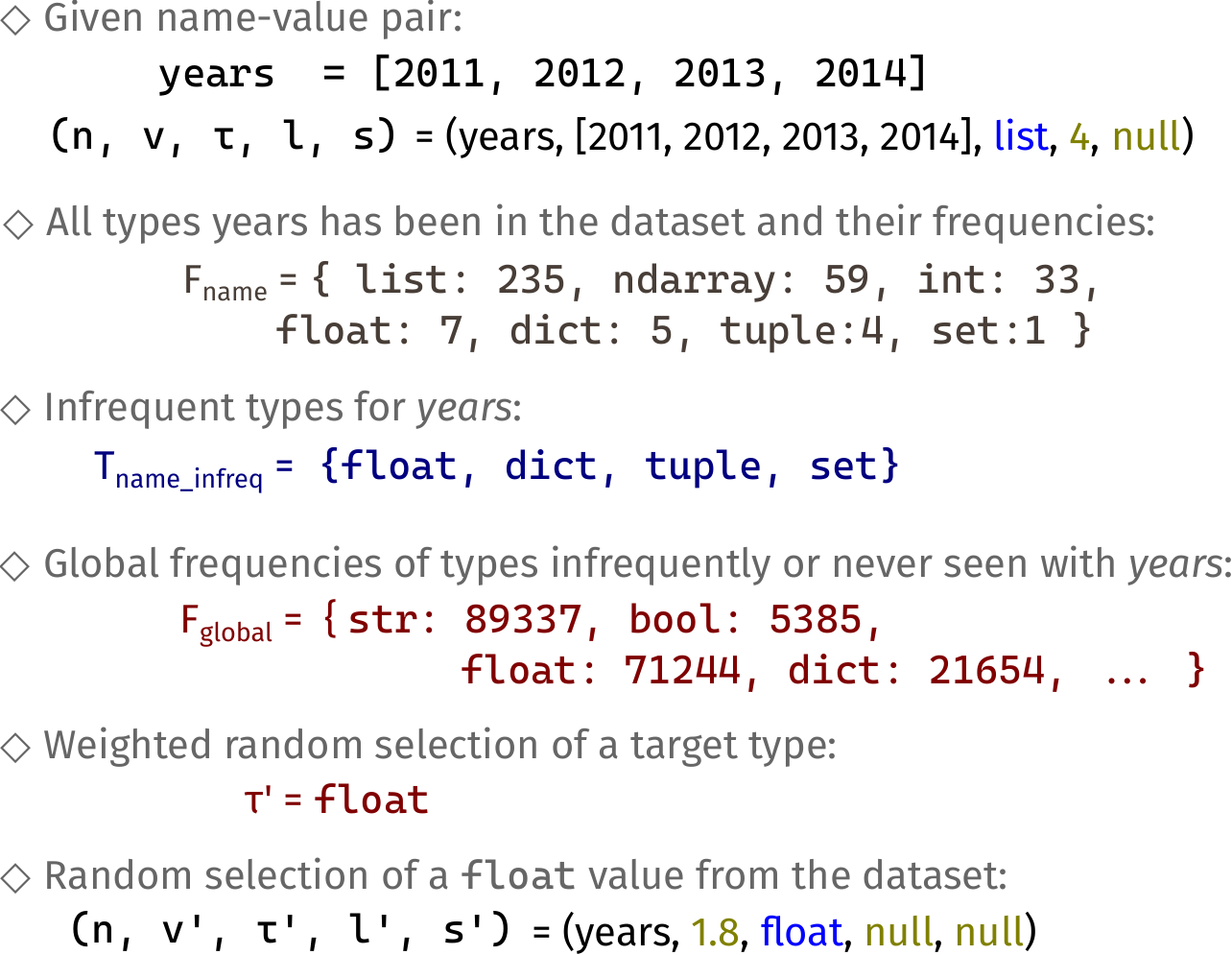}
    \caption{Steps for creating a negative example.}
    \label{fig:example_negative}
\end{figure}

Figure~\ref{fig:example_negative} illustrates the algorithm with an example from our evaluation.
The goal is to create a negative example for a name-value pair where the name is \code{years}.
In the dataset of positive examples, the name \code{years} occurs with values of types \emph{list}, \emph{ndarray}, \emph{int}, etc., with the frequencies shown in the figure.
For example, \code{years} occurs 235 times with a \emph{list} value, but only seven times with a \emph{float} value.
Among all types that occur in the dataset, many never occur together with the name \code{years}, e.g., \emph{str} and \emph{bool}.
Based on the global frequencies of types that \code{years} never or only infrequently occurs with, the algorithm picks \emph{float} as the target type.
Finally, a corresponding \emph{float} value is selected from the dataset, which is \textit{1.8} for the example, and the negative example shown at the bottom of the figure is returned.

By default, \name{} uses the type-guided generation of negative examples, and our evaluation compares it with the purely random technique.
The generated negative examples are combined with the positive examples in the dataset, and the joint dataset serves as training data for the neural classifier.
Due to the automated generation, a generated negative examples may coincidentally be identical to an existing positive example.
In practice, the dataset used during the evaluation contains 38 instances of identical positive and negative examples out of 490,332 negative examples.

\subsection{Representation as Vectors}

\begin{figure}
    \includegraphics[width=.9\linewidth]{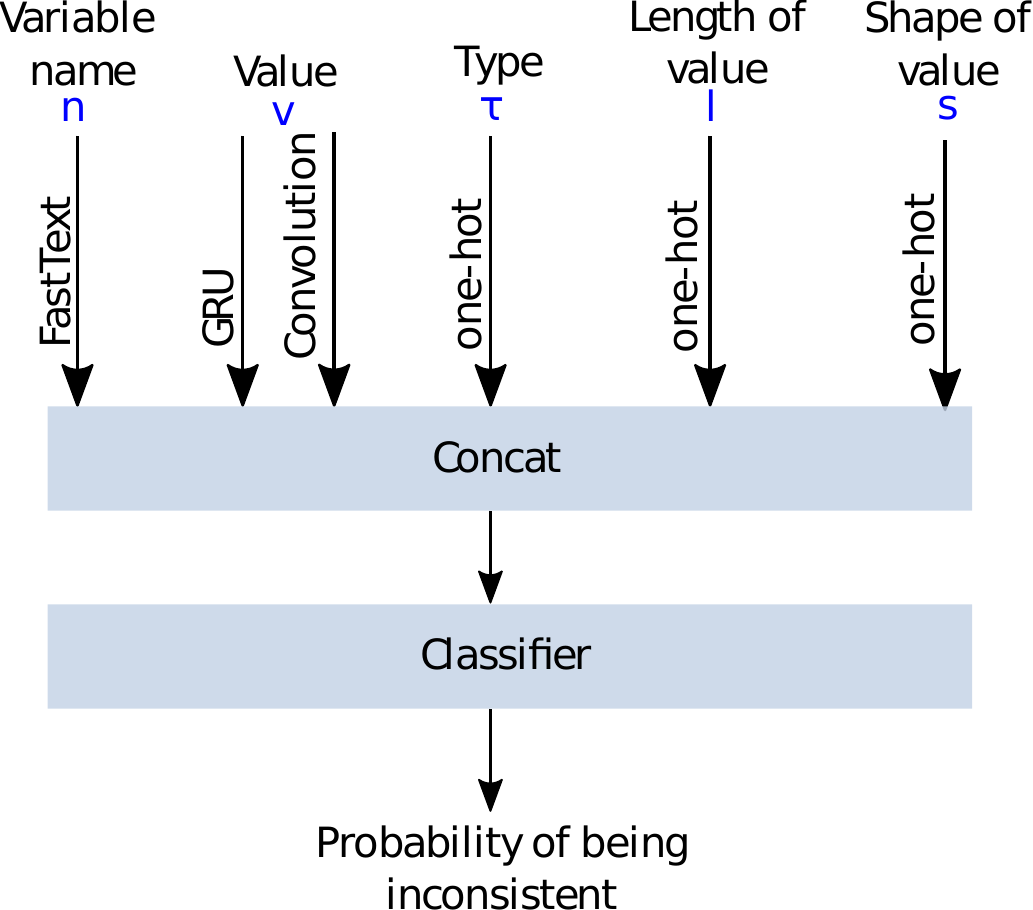}
    \caption{Architecture of the neural model.}
    \label{fig:model}
\end{figure}

Given a dataset of name-value pairs, each labeled either as a positive or a negative example, \name{} trains a neural classification model to distinguish the two kinds of examples.
A crucial step is to represent the information in a name-value pair as vectors, which we explain in the following.
The approach first represents each of the five components $(n, v, \tau, l, s)$ of a name-value pair as a vector, and then feeds the concatenation of these vectors into the classifier.
Figure~\ref{fig:model} shows an overview of the neural architecture.
The following describes the vector representation in more detail, followed by a description of the classifier in Section~\ref{sec:training_and_prediction}.

\paragraph{Representing Variable Names}
To enable \name{} to reason about the meaning of variable names, it maps each name into a vector representation that encodes the semantics of the name.
For example, the representation should map the names \code{list\_of\_numbers} and \code{integers} to similar vectors, as both represent similar concepts, but the vector representations of the names \code{age} and \code{file\_name} should differ from the previous vectors.
To this end, our approach builds on pre-trained word embeddings, i.e., a learned function that maps each name into a vector.
Originally proposed in natural language processing as a means to represents words~\cite{Mikolov2013a,Bojanowski2017}, word embeddings are becoming increasingly popular also on source code~\cite{icse2019,oopsla2018-DeepBugs,Alon2018,DBLP:conf/acl/ChangPCR18,Nguyen2017}, where they represent individual tokens, e.g., variable names.

We build upon FastText~\cite{Bojanowski2017}, a neural word embedding known to represent the semantics of identifiers more accurately than other popular embeddings~\cite{icse2021}.
An additional key benefit of FastText is to avoid the out-of-vocabulary problem that other embeddings, e.g., Word2vec~\cite{Mikolov2013a} suffer from, by splitting each token into n-grams and by computing a separate vector representation for each n-gram.
To obtain meaningful embeddings for the Python domain, we pre-train a FastText model on token sequences extracted from the corpus Python programs used in our evaluation.
Formally, the trained FastText model $M$, assigns to each name $n$ a real-valued vector $M(n) \in \mathbb{R}^d$, where $d=100$ in our evaluation.

\paragraph{Representing Values}
The key challenge for representing the string representations of values as vectors is that there is a wide range of different values, including sequential structures, e.g., in values of types \textit{string}, \textit{ndarray}, \textit{list}, and values without an obvious sequential structure, e.g., primitives and custom objects.
The string representations of values may capture many interesting properties, including and beyond the information conveyed by the type of a value.
For example, the string representation of an \emph{int} implicitly encodes whether the value is a positive or negative number.
Our goal when representing values as vector is to pick up such intricacies, without manually defining type-specific vector encoders. 

To this end, \name{} represents value as a combination of two vector representations, each computed by a neural model that we jointly learn along with the overall classification model.
On the one hand, we use a recurrent neural network (RNN) suitable for capturing sequential structures.
Specifically, we apply gated recurrent units (GRU) over the sequence of characters, where each character is used as an input at every timestep.
The vector obtained from the hidden state of the last timestep then serves as the representation of the complete sequence.
On the other hand, we use a convolutional neural network (CNN) suitable for capturing non-sequential information about the value.
Specifically, the approach applies a one-dimensional CNN over the sequence of characters, where the number of channels for the CNN is equal to the number of characters in the string representation of the value, the number of output channels is set to 100, Relu is the activation function, and a one-dimensional MaxPool layer serves as the final layer.
Finally, \name{} concatenates the vectors obtained from the RNN and the CNN into the overall vector representation of the value.

\paragraph{Representing Types}
To represent the type of a value as a vector, the approach computes a one-hot vector for each type. Each vector has a dimension equal to the number of types present in the dataset.
A type is represented by setting an element to one while keeping the remaining elements set to zero. For example, if we have only three types namely \textit{int, float,} and \textit{list} in our dataset then using one-hot encoding, each of them can be represented as \textit{[1, 0, 0], [0, 1, 0]} and \textit{[0, 0, 1]} respectively.  For the evaluation, we set the maximum number of types to ten. More sophisticated representations of types, e.g., learned jointly with the overall model~\cite{Allamanis2020}, could be integrated into \name{} as part of future work.

\paragraph{Representing Length and Shape}
Length and shape are similar concepts, and hence, we represent them in a similar fashion.
%
Because the length of a value is theoretically unbounded, we consider ten ranges of lengths and represent each of them with a one-hot vector.
Specifically, \name{} considers ranges of length 100, starting from 0 until 1,000.
That is, any length between 0 and 100 will be represented by the same one-hot vector, and likewise any length greater than 1,000 will be represented by the another vector.
The shape of a value is a tuple of discrete numbers, which we represent similarly to the length, except that we first multiply the elements of the shape tuple.
For example, for a value of shape $x,y,z$, we encode $x \cdot y \cdot z$ using the same approach as for the length.
For values that do not have a length or shape, we use a special one-hot vector.

\subsection{Training and Prediction}
\label{sec:training_and_prediction}

Once \name{} has obtained a vector representation for each component of a name-value pair, the individual vectors are concatenated into the combined representation of the pair.
We then feed this combined representation into a neural classifier that predicts the probability $p$ of the name-value pair to be inconsistent.
The classification model consists of two linear layers with a sigmoid activation function at the end.
We also add a dropout with probability of 0.5 before each linear layer.
We train the model with a batch size of 128, using the Adam~\cite{kingma2014adam} optimizer, for 15 epochs, after which the validation accuracy saturates.
During training, the model is trained toward predicting $p=0.0$ for all positive examples and $p=1.0$ for all negative examples.
Once trained, we interpret the predicted probability $p$ as the confidence \name{} has in flagging a name-value pair as inconsistent, and the approach reports to the user only pairs with $p$ above some threshold (Section~\ref{sec:eval effectiveness}).

\subsection{Heuristic Filtering of Likely False Positives}
\label{sec:heuristics}

Before reporting name-value pairs that the model predicts as inconsistent to the user, \name{} applies two simple heuristics to prune likely false positives.
The heuristics aim at removing generic and meaningless names that have passed the filtering described in Section~\ref{sec:data filtering}, such as \code{data} and \code{val\_0}.
The rationale is that judging whether those names match a specific value is difficult, but the goal of \name{} is to identify name-value pairs that clearly mismatch.
The first heuristic removes pairs with names that contain one of the following terms, which are often found in generic names: \code{data}, \code{value}, \code{result}, \code{temp}, \code{tmp}, \code{str}, and \code{sample}.
The second heuristic removes pairs with short and cryptic names.
To this end, we tokenize names at underscores and then remove pairs with names where at least one subtoken has less than three characters.

%
%
\section{Evaluation}
\label{sec:evaluation}

Our evaluation focuses on the following research questions:

\begin{itemize}
    \item RQ1: How effective is the neural model of \name{} in detecting name-value inconsistencies?
    \item RQ2: Are the inconsistencies that \name{} reports perceived as hard to understand by software developers?
    \item RQ3: What kinds of inconsistencies does the approach find in real-world code?
    \item RQ4: How does our approach compare to popular static code analysis tools?
    \item RQ5: How does \name{} compare to simpler variants of the approach?
\end{itemize}

\subsection{Experimental Setup}
\label{subsec:exprimental_setup}

We implement our approach for Python as it is one of the most popular dynamically typed programming languages\cite{Tiobe}.
All experiments are run on a machine with Intel Xeon E5-2650 CPU having 48 cores, 64GB of memory and an NVIDIA Tesla P100 GPU. The machine runs Ubuntu 18.04, and we use Python 3.8 for the implementation. 

The evaluation requires a large-scale, diverse, and realistic dataset of closed (i.e., include all inputs) programs.
We choose one million computational notebooks in an existing dataset of Jupyter notebooks scrapped from GitHub~\cite{10.1145/3173574.3173606}.
The dataset is (i) large-scale because there are many notebooks available,
(ii) diverse because they are written by various developers and cover various application domains,
(iii) realistic because Jupyter notebooks are one of the most popular ways of written Python code these days, and (iv) closed because notebooks do not rely on user input.
Another option would be to apply \name{} to executions of test suites, which often focus on unusual inputs though and, by definition, exercise well-tested and hence likely correct behavior.

Excluding some malformed notebooks, we convert 985,865 notebooks into Python scripts using \textit{nbconvert}.
Some of these notebooks contain only text and no code, while for others, the code has syntax errors, or the code is very short and does not perform any assignments.
All of this decreases the number of Python files that \name{} can instrument, and we finally obtain 598,321 instrumented files.
The instrumentation takes approximately two hours.

When gathering name-value pairs, we face general challenges related to reproducing Jupyter notebooks~\cite{wang2020assessing}.
First, even with the installation of the 100 most popular Python packages, unresolved dependencies result in crashes during some executions.
Second, some Python scripts read inputs from files, e.g., a dataset for training a machine learning model, which may not be locally available.
Considering all notebooks that we can successfully execute despite these obstacles, \name{} gathers a total of 947,702 name-value pairs, of which 500,332 remain after the filtering described in Section~\ref{sec:data filtering}.
The extracted pairs come from 106,652 Python files with a total of 7,231,218 lines of non-comment, non-blank Python code.
Running the instrumented files to extract name-value pairs takes approximately 48 hours.

Before running any experiments with the model, we sample 10,000 name-value pairs as a held-out \textit{test dataset}. Unless mentioned otherwise, all reported results are on this test dataset.
On the remaining 490,332 name-value pairs, we perform an 80-20 split into \textit{training} and \textit{validation} data.
For each name-value pair present in the training, validation, and test datasets, we create a corresponding negative example, which takes two hours in total.
The total number of data points used to train the \name{} model hence is about 780k. Training takes an average of 190 seconds per epoch and once trained, prediction on the entire test dataset takes about 15 seconds.

We find the name-value pairs to consist of a diverse set of values and types.
There are 99.8k unique names, i.e., each name appears, on average, about 10 times.
The top-5 frequent types are 
\emph{list}, \emph{ndarray}, \emph{str}, \emph{int}, \emph{float}. 
The presence of a large number of collection types, such as \emph{list} and \emph{ndarray}, which usually are not fully initialized as literals shows that extracting values at run-time is worthwhile.



\subsection{RQ1: Effectiveness of the Trained Model}
\label{sec:eval effectiveness}

We measure the effectiveness of \name{}'s model by applying the trained model to the held-out test dataset.
The output of the model can be interpreted as a confidence score that indicates how likely the model believes a given name-value pair to be inconsistent.
We consider all name-value pairs $P_{\mathit{warning}}$ with a score above some threshold as a warning, and then measure precision and recall of the model w.r.t.\ the inconsistency labels in the dataset ($P_{inconsistent}$ are pairs labeled as inconsistent):
\begin{center}
$\mathit{precision} = \frac{|P_{\mathit{warning}} \cap P_{\mathit{inconsistent}}|}{|P_{\mathit{warning}}|}$\\
$\mathit{recall} = \frac{|P_{\mathit{warning}} \cap P_{\mathit{inconsistent}}|}{|P_{\mathit{inconsistent}}|}$
\end{center}
We also compute the F1 score, which is the harmonic mean of precision and recall.

\begin{figure}
    \includegraphics[width=.9\linewidth]{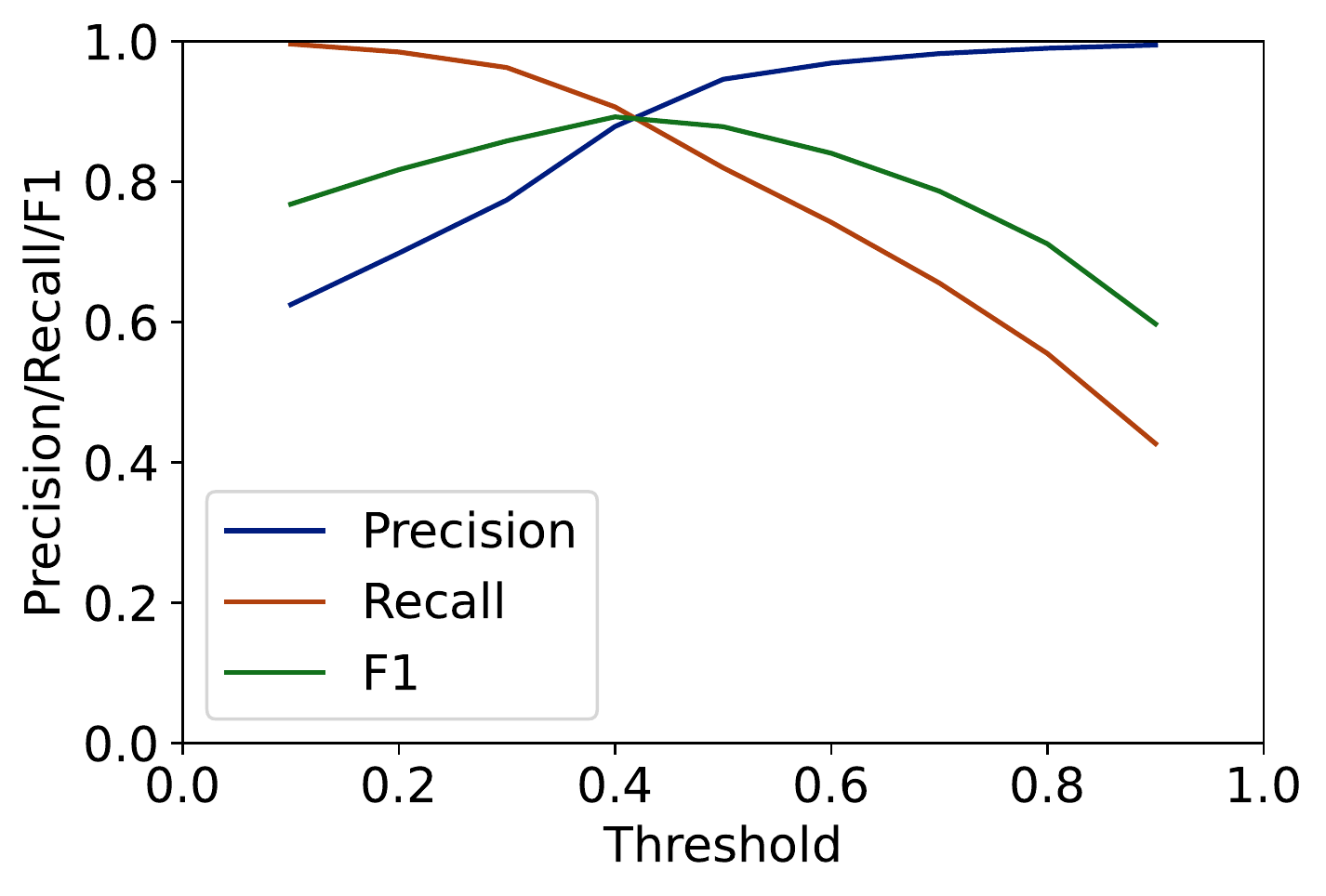}
    \caption{Precision, recall, and F1 score with different thresholds for reporting warnings.}
    \label{fig:p_r_curve}
\end{figure}

Figure~\ref{fig:p_r_curve} shows the results for different thresholds for reporting a prediction as a 
warning.
The results illustrate the usual precision-recall tradeoff, where a user can reduce the risk of false positive warnings at the cost of finding fewer inconsistencies.
The model achieves the highest F1 score of 89\% at a threshold of 0.4, with a precision of 88\% and a recall of 91\%.
Unless otherwise mentioned, we use a threshold of 0.5 as the default, which gives 87\% F1 score.
Out of 8,858 files in the held-out test set, 336 (3.8\%) have at least one warning reported by \name{}.

\begin{finding}
The model effectively identifies inconsistent name-value pairs, with a maximum F1 score of 89\%.
\end{finding}

\subsection{RQ2: Study with Developers}
To answer the question how well \name{}'s warnings match name-value pairs that developers perceive as hard to understand, we perform a study with eleven software developers.
The participants are four PhD students and seven master-level students, all of which regularly develop software, and none of which overlaps with the authors of this paper.
During the study, each participant is shown 40 name-value pairs and asked to assess each pair regarding its understandability.
The participants provide their assessment on a five-point Likert scale ranging from ``hard''~(1) to ``easy''~(5), where ``hard'' means that the name and the value are inconsistent, making it hard to understand and maintain the code.
The 40 name-value pairs consist of 20 pairs that are randomly selected from all warnings \name{} reports as inconsistent with a confidence above 80\% and of 20 randomly selected pairs that the approach does not warn about.
For each pair, the participants are shown the name of the variable, the value that \name{} deems inconsistent with this name, and the type of the value.
In total, the study hence involves 440 developer ratings.
Because what is a meaningful variable names is, to some extent, subjective, we expect some variance in the ratings provided by the participants.
To quantify this variance, we compute the inter-rater agreement using Krippendorff's alpha, which yields an agreement of 56\%.
That is, developers agree with a medium to high degree on whether a name-value pair is easy to understand.

Before providing quantitative results, we discuss a few representative examples.
Among the name-value pairs without a warning is a variable called \code{DATA\_URL} that stores a string containing a URL.
This pair is consistently rated as easy to understand, with a mean ranking of 5.0.
Among the pairs that \name{} reports as inconsistent are a variable \code{password\_text} storing an integer value \code{0}, which most participants consider as hard to understand (mean rating: 1.54).
Another pair that the approach warns about is a variable called \code{path} that stores an empty list.
The study participants are rather undecided about this example, with a mean rating of 2.72.


\begin{figure}
	\begin{subfigure}[b]{\linewidth}
         \centering
         \begin{tabular}{@{}l|rr@{}}
         \toprule
         & \multicolumn{2}{c}{\name{}'s prediction} \\
         \cmidrule{2-3}
         Developer & Consistent  & Inconsistent \\
         assessment & ($P_{\mathit{noWarning}}$) & ($P_{\mathit{warning}}$) \\
         \midrule
         Easy to understand ($P_{\mathit{easy}}$) & 15 & 4 \\
         Hard to understand ($P_{\mathit{hard}}$) & 5 & 16 \\
         \bottomrule
         \end{tabular}
         \caption{\name{} predictions of inconsistencies vs.\ developer-perceived understandability.}
         \label{fig:per pair comparison}
     \end{subfigure}
 	 \begin{subfigure}[b]{\linewidth}
         \centering
         \includegraphics[width=.9\linewidth]{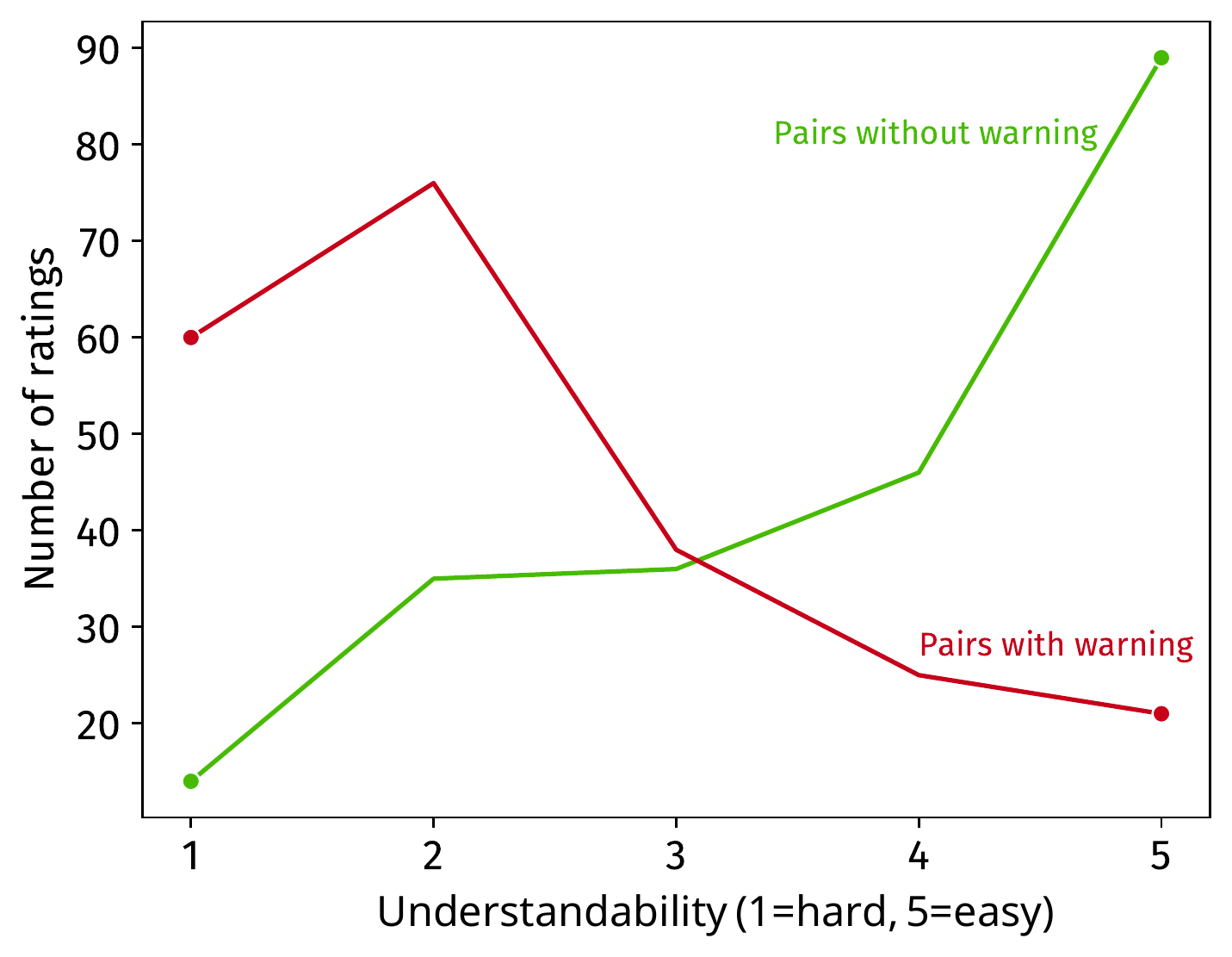}
         \caption{Understandability ratings for name-value pairs with and without warnings by \name{}.}
         \label{fig:all ratings}
     \end{subfigure}
	\caption{Results from user study.}
	\label{fig:user study}
\end{figure}

\begin{table*}[t]
	\centering
  \caption{Examples of warnings produced by~\name{}.}
	\label{tab:examples_of_warnings_category}
  \small
\setlength{\tabcolsep}{7pt}
    \begin{tabular}{@{}p{23em}p{4em}p{8em}p{23em}@{}} \toprule
        Code Example & Category & Run-time value & Comment \\
\midrule
\vspace{-1.2em}
\begin{lstlisting}[escapechar=§]
name = 'Philip K. Dick'
...
§\colorbox{hlyellow}{\makebox(30,3){\strut\textcolor{black}{name = \textcolor{orange1}{2.5}}}}§
if type(name) == str:
  print('yes')
\end{lstlisting}
 &
\vspace{-1em}
Misleading name
&
\code{2.5}
&
\vspace{-1em}
A variable called \code{name} is typically holding a string, but here stores a float value.
\\ 
\midrule
\vspace{-1.2em}
\begin{lstlisting}[escapechar=§]
§\colorbox{hlyellow}{\makebox(130,3){\strut\textcolor{black}{file = os.path.exists(\textcolor{green}{'reference.csv'})}}}§
if file == False: 
     print('Warning: ...')
\end{lstlisting}
 &
\vspace{-1em}
Misleading name
&
\code{False}
&
\vspace{-1em}
The name \code{file} suggests that the variable stores either a file handle or a file name, but it here stores a boolean.
\\ 
\midrule
\vspace{-1.2em}
\begin{lstlisting}[escapechar=§]
def Custom(information):
    §\colorbox{hlyellow}{\makebox(155,3){\strut\textcolor{black}{prob =  \textcolor{cyan}{get\_betraying\_probability}(information)}}}§
    if(prob > 1 / 2):
      return D
    elif(prob == 1 / 2):
      return choice([D, C])
    else:
      return C
\end{lstlisting}
&
\vspace{-1em}
Incorrect value
&
\code{"Corporate"}
&
\vspace{-1em}
Assigning a string to a variable called \code{prob} is unusual, because \code{prob} usually refers to a probability. The value is incorrect and leads to a crash in the next line because comparing a string and a float causes a type error.
\\ 
\midrule
\vspace{-1.2em}
\begin{lstlisting}[escapechar=§]
§\colorbox{hlyellow}{\makebox(179,3){\strut\textcolor{black}{dwarF = \textcolor{green}{'/Users/iayork/Downloads/dwar\_2013\_2015.txt'}}}}§
dwar = pd.read_csv(dwarF, sep=' ', header=None)
\end{lstlisting}
 &
\vspace{-1em}
False positive
&
\code{"/Users/.."}
&
\vspace{-1em}
The value is a string that describes file path, which fits the name, where the \code{F} supposedly means ``file''. The model reports this false positive because it fails to understand the abbreviation.
\\ 
\bottomrule
\end{tabular}
\end{table*}

The main question of the user study is to what extent \name{} pinpoints name-value pairs that developers also consider to be hard to understand.
We address this question in two ways, first by computing precision and recall of \name{} w.r.t. the developer ratings, and then by comparing the ratings for warnings and non-warnings.

\paragraph{Precision and Recall w.r.t. Developer Ratings}
We assign each of the 40 name-value pairs into two sets:
On the one hand, a pair is in $P_{\mathit{hard}}$ if the mean rating assigned by the developers is less than three and in $P_{\mathit{easy}}$ otherwise.
On the other hand, a pair is in $P_{\mathit{warning}}$ if \name{} flags it as an inconsistency and in $P_{\mathit{noWarning}}$ otherwise.
Table~\ref{fig:per pair comparison} shows the intersections between these sets.
For example, we see that 16 of the pairs that \name{} warns about, but only 5 of the pairs without a warning, are considered to be hard to understand.
We compute precision and recall as follows:
\vspace{-.1em}
\begin{center}
$\mathit{precision} = \frac{|P_{\mathit{warning}} \cap P_{\mathit{hard}}|}{|P_{\mathit{warning}}|} = \frac{16}{20} = 80\%$\\
$\mathit{recall} = \frac{|P_{\mathit{warning}} \cap P_{\mathit{hard}}|}{|P_{\mathit{hard}}|} = \frac{16}{21} = 76\%$
\end{center}
\paragraph{Ratings for Warnings vs. Non-Warnings}
In addition to the pair-based metrics above, we also globally compare the ratings for pairs with and without warnings.
The goal is to understand whether \name{} is effective at distinguishing between name-value pairs that developers perceive as easy and hard to understand.
To this end, consider two sets of ratings:
ratings $R_{\mathit{warning}}$ for name-value pairs that \name{} reports as inconsistent, and
ratings $R_{\mathit{noWarning}}$ for other name-value pairs.
Figure~\ref{fig:all ratings} compares the two sets of ratings with each other, showing how many ratings there are for each point on the 5-point Likert scale.
The results show a clear difference between the two sets: ``easy'' is the most common rating in $R_{\mathit{noWarning}}$, whereas the majority of ratings in $R_{\mathit{warning}}$ is either ''relatively hard'' or ``hard''.
We also statistically compare $R_{\mathit{warning}}$ and $R_{\mathit{noWarning}}$ using a Mann-Whitney U-test, which shows the two sets of rankings to be extremely likely to be sampled from different populations (with a p-value of less than 0.1\%). 

\begin{finding}
Developers mostly agree with the (in)consistency predictions by \name{}. In particular, they assess 80\% of the name-value pairs that the approach warns about as hard to maintain and understand.
\end{finding}

\subsection{RQ3: Kinds of Inconsistencies in Real-World Code}
\label{subsec:rq2-kinds_of_inconsistencies}

To better understand the kinds of name-value inconsistencies detected in real-world code, we inspect name-value pairs in the test datasets that appear as such in the code, but that are classified as inconsistent by the model.
When using \name{} to search for previously unknown issues, these name-value pairs will be reported as warnings.
We inspect the top-30 predictions, sorted by the probability score provided by the model, and classify each warning into one of three categories:
\begin{itemize}
    \item \textit{Misleading name}. Name-value pairs where the name clearly fails to match the value it refers to. These cases do not lead to wrong program behavior, but should be fixed to increase the readability and maintainability of the code.
    \item \textit{Incorrect value}. Name-value pairs where the mismatch between a name and a value is due to an incorrect value being assigned. These cases cause unexpected program behavior, e.g., a program crash or incorrect output.
\item \textit{False positive}. Name-value pairs that are consistent with each other, and which ideally would not be reported as a warning.
\end{itemize}

The inspection shows that 21 of the warnings correspond to misleading names, 2 are incorrect values, and 7 are false positives.
That is, the majority of the reported inconsistencies are due to the name, whereas only a few are caused by an incorrect value being assigned to a meaningful name.
This result is expected because incorrect behavior is easier to detect, e.g. via testing, than misleading names, for which currently few tools exist.
The fact that 23 out of 30 warnings (77\%) are true positives is also consistent with the developer study in RQ2.

Table~\ref{tab:examples_of_warnings_category} shows representative examples of warnings produced by \name{}.
The first two examples show misleading names.
For example, it is highly unusual to assign a number to a variable called~\code{name} or to assign boolean to a variable called \code{file}.
To the best of our knowledge, these misleading names do not cause unexpected behavior, but developers may still want to fix them to increase the readability and maintainability of the code.
In the third example, \name{} produces a warning about the assignment on line 2. The value assigned during the execution is a string \code{'Cooperate'}. Due to the string assignment, the code on line 3 crashes since the operator \code{>} does not support a comparison between a string and float.
\name{} is correct in predicting this warning because the variable name \code{prob} is typically used to refer to a probability, not to a string like \code{'Cooperate'}.
The final example is a false positive, which illustrates one of the most common causes of false positives seen during our inspection, namely short (and somewhat cryptic) names for which the model fails to understand the meaning.




\begin{finding}
The majority of inconsistencies detected in real-world code are due to the name in a name-value pair being misleading, and occasionally also due to incorrect values.
\end{finding}

     

\subsection{RQ4: Comparison with Previous Bug Detection Approaches}
We compare \name{} to three state-of-the-art static analysis tools aimed at finding bugs and other kinds of noteworthy issues:
(i) \textit{pyre}, a static type checker for Python that infers types and uses available type annotations. We compare with pyre because many of the inconsistencies that \name{} reports are type-related, and hence, might also be spotted by a type checker.
(ii) \textit{flake8}, a Python linter that warns about commonly made mistakes. We compare with flake8 because it is widely used and because linters share the goal of improving the quality of code.
(iii) \textit{DeepBugs}~\cite{oopsla2018-DeepBugs}, a learning-based bug detection technique. We compare with DeepBugs because it also aims to find name-related bugs using machine learning, but using static instead of dynamic analysis.
We run pyre and flake8 using their default configurations.
For DeepBugs, we install the ``DeepBugs for Python'' plugin from the marketplace of the PyCharm IDE.
We apply each of the three approaches to the 30 files where \name{} has produced a warning and which have been manually inspected (RQ3).
Namer~\cite{He2021}, a recent technique for finding name-related coding issues through a combination of static analysis, pattern mining, and supervised learning would be another candidate for comparing with, but neither the implementation nor the experimental results are publicly available.

\begin{table}[t]
  \centering
  \caption{Comparison with existing static bug detectors.}
  \label{tab:comparison_with_static_analysis_and_deepbugs}
  \small
  \setlength{\tabcolsep}{18pt}
  \begin{tabular}{@{}lrr@{}}
    \toprule
    Approach                      & Warnings         & Warnings common with \name{} \\
    \midrule
    pyre & 54 & 1/30 \\
    flake8 & 1,247 & 0/30 \\
    DeepBugs & 151 & 0/30 \\
    \bottomrule
  \end{tabular}
\end{table}

Table~\ref{tab:comparison_with_static_analysis_and_deepbugs} shows the number of warnings reported by the existing tools and how many of these warnings overlap with those reported by \name{}.
We find that except one warning reported by pyre, none matches with the 30 manually inspected warnings from \name{}.
The matching warning is a misleading name, shown on the first row of Table~\ref{tab:examples_of_warnings_category}.
The pyre type checker reports this as an  ``Incompatible variable type'' because in the same file, the variable \scode{name} is first assigned a string \scode{'Philip K. Dick'} and later assigned a float value \scode{2.5}. 
The 1,247 warnings produced by flake8 are mostly about coding style, e.g., ``missing white space'' and ``whitespace after '(' ''.
The warnings reported by DeepBugs include possibly wrong operator usages and incorrectly ordered function arguments, but none matches the warnings reported by \name{}.

\begin{finding}
\name{} is complementary to both traditional static analysis-based tools and to a state-of-the-art learning-based bug detector aimed at name-related bugs.
\end{finding}

\subsection{RQ5: Comparison with Variants of the Approach}

\subsubsection{Type-Guided vs.\ Purely Random Negative Examples}
The following compares the two algorithms for generating negative examples described in Section~\ref{subsec:create_negative_examples}.
Following the setup from RQ1, we find that the purely random generation reduces both precision and recall, leading to a maximum F1 score of 0.82, compared to 0.89 with the type-guided approach.
Manually inspecting the top-30 reported warnings as in RQ2, we find 21 false positives, nine misleading names, and zero incorrect values, which clearly reduces the precision compared to the type-guided generation approach.
These results confirm motivation for the type-guided algorithm (Section~\ref{sec:type-guided}) and show that it outperforms a simpler baseline.

\subsubsection{Ablation Study}

\begin{figure}
    \includegraphics[width=.48\textwidth]{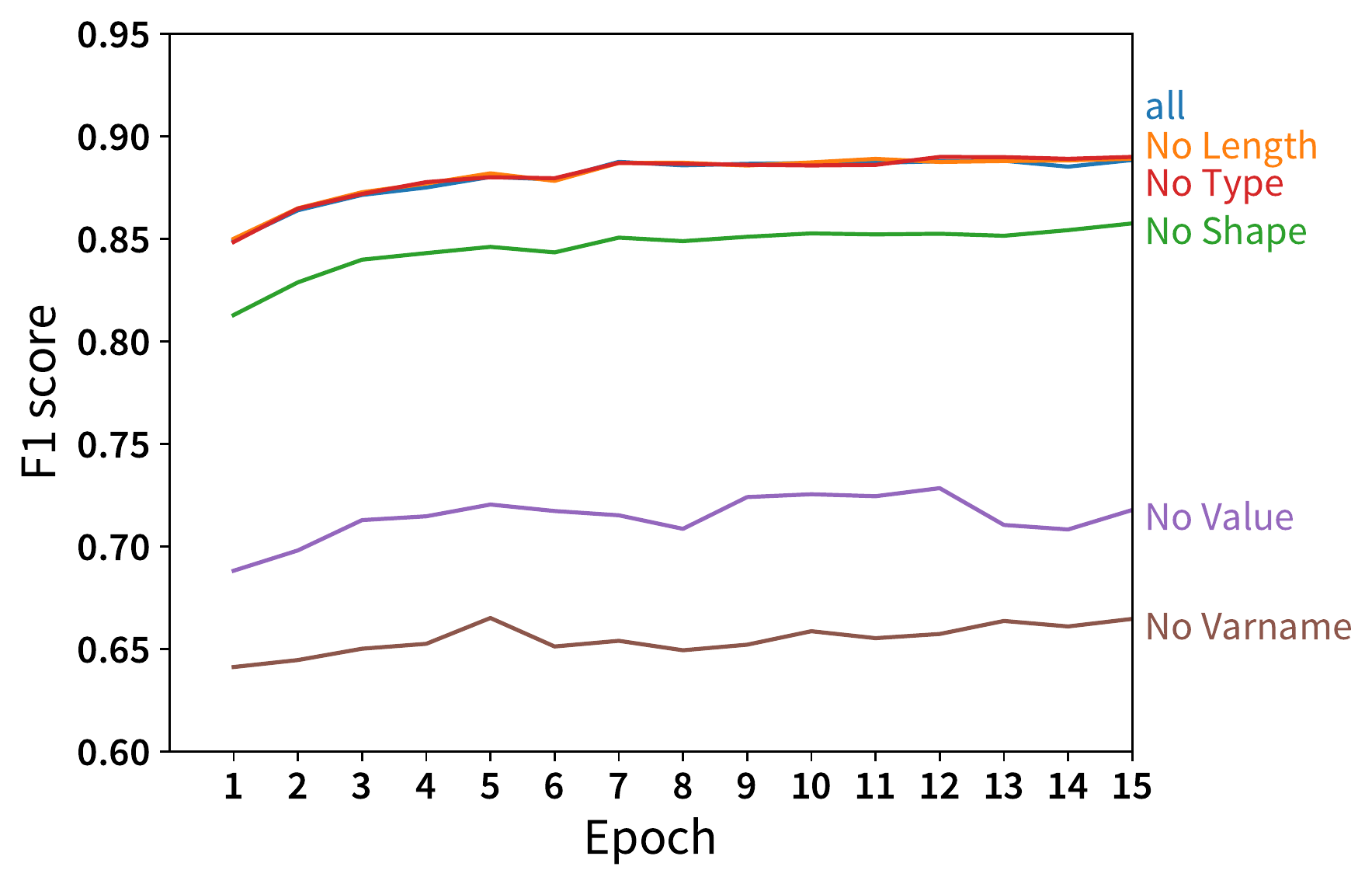}
    \caption{Result of ablation study.}
    \label{fig:ablation_study_results}
\end{figure}

We perform an ablation study to measure the importance of the different components of a name-value pair fed into the model.
To this end, we set the vector representation of individual components to zero during training and prediction, and then measure the effect on the F1 score of the model.
Figure~\ref{fig:ablation_study_results} shows the results, where the vertical axis shows the F1 score obtained on the validation dataset at each epoch during training.
Each line in Figure~\ref{fig:ablation_study_results} shows the F1 score obtained while training the model keeping that particular feature set to zero. For example, the green line (``No Shape'') is for a model that does not use the shape of a value, and the blue line (``all'') is for a model that uses all components of a name-value pair.
We find that the most important inputs to the model are the variable name and the string representation of the value.
Removing the length or the type of a value does not significantly decrease the model's effectiveness. The reason is that these properties can often be inferred from other inputs given to the model, e.g., by deriving the type from the string representation of a value.
We confirm this explanation by removing both the type and the string representation of a value, which yields an F1 score similar to the model trained by removing only values.

\begin{finding}
Each component of the approach contributes to the overall effectiveness, but there is some redundancy in the properties of values given to the model.
\end{finding}

\section{Threats to Validity}

\paragraph{Internal Validity}
Several factors may influence our results.
First, the filtering of name-value pairs based on the length of names may accidentally remove short but meaningful names, such as abbreviations that are common in a specific domain.
Preliminary experiments without such filtering resulted in many false positives, and we prefer false negatives over false positives to increase developer acceptance.
Second, our manual classification into different kinds of inconsistencies is subject to our limited knowledge of the analyzed Python files.
To mitigate this threat, the classification is done by two of the authors, discussing any unclear cases until reaching consensus.

\paragraph{External Validity}
Several factors may influence the generalizability of our results.
First, our approach is designed with a dynamically typed programming language in mind, because meaningful identifier names are particularly important in such languages.
This focus and the setup of our experiments implies that we cannot draw conclusions beyond Python or beyond the kind of Python code found in Jupyter notebooks.
Second, our developer study is limited to eleven participants, and other developers may assess the understandability of the name-value pairs differently.
We mitigate this threat by getting eleven opinions about each name-value pair and by statistically analyzing the relevance of the results.


\section{Related Work}

\paragraph{Detecting Poor Names}
The importance of meaningful names during programming has been studied and established~\cite{Lawrie2006,Butler2010}.
There are several techniques for finding poorly named program elements, e.g.,
based on pre-defined rules~\cite{abebe2009lexicon},
by comparing method names against method bodies~\cite{Host2009}, and
through a type inference-like analysis of names and their occurrences~\cite{Lawall2010}.
To improve identifier names, rule-based expansion~\cite{lawrie2011expanding}, n-gram models of code~\cite{Allamanis2014}, and learning-based techniques that compare method bodies and method names have been proposed~\cite{Liu2019,Nguyen2020}.
Namer~\cite{He2021} combines static analysis, pattern mining, and supervised learning to find name-related coding issues.
Many of the above approaches focus on method names, whereas we target variables.
Moreover, none of the existing approaches exploits dynamically observed values.

\paragraph{Predicting Names} 
When names are completely missing, e.g., in minified, compiled, or obfuscated code, learned models can predict them~\cite{Raychev2015,Vasilescu2017,Context2Name,Lacomis2019}.
Another line of work predicts method names given the body of a method~\cite{Allamanis2015,Allamanis2016,Alon2018}, which beyond being potentially useful for developers serves as a pseudo-task to force a model to summarize code in a semantics-preserving way.
\name{} differs by considering values observed at runtime, and not only static source code, and by checking names for inconsistencies with the values they refer to, instead of predicting names from scratch.

\paragraph{Name-based Program Analysis}
DeepBugs introduced learning-based and name-based bug detection~\cite{oopsla2018-DeepBugs}, which differs from \name{} by being purely static and by focusing on different kinds of errors.
The perhaps most popular kind of name-based analysis is probabilistic type inference~\cite{Xu2016}, often using deep neural network models~\cite{Hellendoorn2018,icse2019,fse2020,Allamanis2020,Wei2020} that reason about the to-be-typed code.
RefiNum uses names to identify conceptual types, which further refine the usual programming language types~\cite{Dash2018}.
SemSeed exploits semantic relations between names to inject realistic bugs~\cite{fse2021}.
All of the above work is based on the observation that the implicit information embedded in identifiers is useful for program analyses.
Our work is the first to exploit this observation to find name-value inconsistencies.

\paragraph{Natural Language vs.\ Code}
Beyond natural language in the form of identifiers, comments and documentation associated with code are another valuable source of information.
iComment~\cite{tan2007icomment} and tComment~\cite{tan2012tcomment} use this information to detect inconsistencies between comments and code.
Our work differs by focusing on variable names instead of comments, by comparing the natural language artifact against runtime values instead of static code, and by using a learning-based approach.
Another line of work uses natural language documentation to infer specifications of code~\cite{pandita2012inferring,Motwani2019,goffi2016automatic}, which is complementary to our work.

\paragraph{Learning on Code}
In addition to the work discussed above, machine learning on code is receiving significant interest recently~\cite{NeuralSoftwareAnalysis}.
Embeddings of code are one important topic, e.g., using AST paths~\cite{Alon2019}, control flow graphs~\cite{Wang2020a}, ASTs~\cite{Zhang2019}, or a combination of token sequences and a graph
representation of code~\cite{Hellendoorn2020}.
Our encoder of variable names could benefit from being combined with an encoding of the code surrounding the assignment using those ideas.
Other work models code changes and then makes predictions about them~\cite{Hoang2020,Brody2020},
or trains models for program repair~\cite{Gupta2017,Dinella2020},
code completion~\cite{DBLP:journals/corr/abs-2004-05249,kim-arxiv-2020,Alon2019a}, and 
code search~\cite{Gu2018,Sachdev2018}.
 
\paragraph{Learning from Executions}
Despite the recent surge of work on learning on code, learning on data gathered during executions is a relatively unexplored area.
One model embeds student programs based on dynamically observed input-output relations~\cite{Piech2015}.
Wang et al.'s ``blended'' code embedding learning~\cite{Wang2020} combines runtime traces, which include values of multiple variables, and static code elements to learn a distributed vector representation of code.
Beyond code embedding, BlankIt~\cite{Porter2020} uses a decision tree model trained on runtime data to predict the library functions that a code location may use.
In contrast to these papers, our work addresses a different problem and feeds one value at a time into the model.

\section{Conclusion}
Using meaningful identifier names is important for code understandability and maintainability.
This paper presents \name{}, which addresses the problem of finding inconsistencies that arise due to the use of a misleading name or due to assigning an incorrect value.
The key novelty of \name{} is to learn from names and their values assigned at runtime.
To reason about the meaning of names and values, the approach embeds them into vector representations that assign similar vectors to similar names and values.
Our evaluation with about 500k name-value pairs gathered from real-world Python programs shows that the model is highly accurate, leading to warnings reported with a precision of 80\% and recall of 76\%.

\medskip
\noindent
Our implementation and experimental results are available:\\
\url{https://github.com/sola-st/Nalin}

\section*{Acknowledgments}
This work was supported by the European Research Council (ERC, grant agreement 851895), and by the German Research Foundation within the ConcSys and Perf4JS projects.


\bibliographystyle{ACM-Reference-Format}
\bibliography{references}

\end{document}